\documentclass[showpacs,aps,pre,twocolumn,floatfix,superscriptaddress]{revtex4}
\usepackage{graphicx,graphics}
\begin {document}
\title{\bf{Compressibility Consideration in the Boundary of a Strongly Collapsing Bubble}}
\author{A. Moshaii}
\email{moshaii@mehr.sharif.edu} \affiliation{Department of
Physics, Sharif University of Technology, P.O. Box:11365-9161,
Tehran, I.R. Iran.} \affiliation{Institute for Studies in
Theoretical Physics and Mathematics, P.O. Box:19395-5531, Tehran,
I.R. Iran.}
\author{R. Sadighi-Bonabi}
\affiliation{Department of Physics, Sharif University of
Technology, P.O. Box:11365-9161, Tehran, I.R. Iran.}
\affiliation{Bonab Research Center, P.O. Box:56515-196, Bonab,
Azarbayejan Province, I.R. Iran.}
\author{M. Taeibi-Rahni}
\affiliation{Department of Aerospace Engineering, Sharif
University of Technology, P.O. Box:11365-9161, Tehran, I.R. Iran}

\begin{abstract}
Equations of radial motion of a gas bubble in a compressible
viscous liquid have been modified to account for compressibility
at the bubble boundary. A new bubble boundary equation has been
derived, including a new term resulted from liquid
compressibility. The influence of this term has been numerically
investigated using isothermal-adiabatic model for the gas inside
the bubble. The results clearly indicate that, at the end of the
collapse, the new term has a very significant role and its
consideration dramatically changes the bubble characteristics.
Moreover, the more intense the collapse is, the more significant
the effect of the new term is. Also, it has been reasoned out
that, the influence of the new term will be established even when
the effects of mass (water vapor) exchange, chemical reactions,
and gas dynamics inside the bubble are taken into account.
\end{abstract}
\maketitle

\section{Introduction}
The problem of non-linear radial oscillations of a gas bubble in a
liquid, when it experiences a high amplitude spherical sound
field, is an old challenging problem. Several complications are
present in this problem arising from the effects of heat
conduction, mass diffusion, compressibility, chemical reactions,
surface tension, and viscosity. Many authors have studied
different aspects of this matter, but a rather complete
description has not been presented yet.

The radial dynamics of a bubble in an incompressible liquid is
described by the well-known incompressible Rayleigh-Plesset
equation \cite{Rayleigh:1917,Noltingk:1951}. The extension of this
equation to the bubble motion in a compressible liquid has been
studied by many authors; for example Herring \cite{Herring:1941},
Trilling \cite{Trilling:1952}, Gilmore \cite{Gilmore:1952},
Keller and Kolodner \cite{Keller:1956}, Flynn \cite{Flynn:1975},
Lastman and Wentzell \cite {Lastman:1979}, L\"{o}fstedt
\textit{et al.} \cite{fstedt:1993}, and Nigmatulin \textit{et
al.} \cite{Nigmatulin:2000}. In these works, several different
forms of the compressible bubble dynamics equations have been
presented, using different approximation for consideration of the
compressibility effects. On the other hand, the influences of heat
conduction and mass diffusion in the bubble motion have been
reported by Hickling \cite{Hickling:1963}, Fujikawa and Akumatsu
\cite{Fujikawa:1980}, and Yasui \cite{Yasui:1997}.

In a generalized approach, Prosperetti and Lezzi
\cite{Prosperetti:1987} used a singular-perturbation method of
the bubble-wall Mach number and derived a one-parameter family of
equations describing the bubble motion in the first order
approximation of compressibility. This family of equations are
written as:
\begin{eqnarray}
\label{eq1} \left({1-(\eta + 1)\frac{\dot{R}}{C}}
\right)\!R\ddot{R}+\frac{3}{2}\left({1-\frac{1}{3}(3\eta +
1)\frac{\dot {R}}{C}}\right)\!\dot{R}^2 = \nonumber \\
{\frac{R}{\rho C}\frac{d}{dt}\left( {P_l-P_a}
\right)\!+\!\left({1+(1-\eta)\frac{\dot {R}}{C}} \
\right)\!\!\!\left( \frac{P_l-P_a-P_0}{\rho}
\right)},\nonumber \\
\end{eqnarray}
\noindent where, $R$, $C$, $P_0$, $P_a$, and $\rho$ are bubble
radius, liquid sound speed, ambient pressure, driving pressure,
and density of the liquid, respectively. Also, $\eta$ is an
arbitrary parameter. Equation (\ref{eq1}) must be supplemented by
a boundary condition equation at the bubble interface to relate
the liquid pressure, $P_{l}$, to the gas pressure inside the
bubble. Like all previous authors, Prosperetti and Lezzi
\cite{Prosperetti:1987} used the following incompressible
equation for this purpose:
\begin{equation}
\label{eq2}
P_l = P_g - 4\mu \frac{\dot {R}}{R}-\frac{2\sigma}{R},
\end{equation}
\noindent where, $P_{g}$, $\mu$, and $\sigma$ are gas pressure at
the bubble interface, liquid first viscosity coefficient, and
surface tension, respectively. As Prosperetti and Lezzi showed
\cite{Prosperetti:1987}, different existing forms of the bubble
dynamics equations belong to this single parameter family of
equations, corresponding to different values of $\eta$.
Especially, $\eta=1$ and $\eta=0$ correspond to Keller and Miksis
equation \cite{Keller:1980} and Herring-Trilling equation
\cite{Herring:1941, Trilling:1952}, respectively.

Two specific approximations have been considered in the derivation
of Eq'ns. (\ref{eq1}) and (\ref{eq2}). First, Eq'n. (\ref{eq1})
has been derived from the Euler equation, in which the liquid
viscosity effects have been neglected. However, these effects are
not important in the usual applications of the bubble dynamics
equations. It can be shown \cite{Moshaii:2002} that these effects
are significant when $\mu \gtrsim \rho C R_0$, where $R_0$ is the
bubble ambient radius. Therefore, for micron size and larger
bubbles the elimination of these effects is thoroughly justified.

The second approximation, which is more important, is the
incompressibility assumption of the liquid and the gas at the
bubble interface, which has been used in the derivation of Eq'n.
(\ref{eq2}). Note that, all of the effects of the liquid compressibility
in the work of Prosperetti and Lezzi, as well as in all previous
works, have been resulted from the liquid motion around the bubble,
but not from the bubble boundary condition equation. In fact, all
previous authors, on one hand take into account the
compressibility of the liquid motion around the bubble, but on
the other hand neglect its consideration at the bubble
interface. Although Eq'n. (\ref{eq2}) has widely been used in all
old and recent works
\cite{Rayleigh:1917,Noltingk:1951,Herring:1941,Trilling:1952,
Gilmore:1952,Keller:1956,Flynn:1975,Lastman:1979,fstedt:1993,
Nigmatulin:2000,Hickling:1963,Fujikawa:1980,Yasui:1997,Prosperetti:1987,Keller:1980},
but its applicability for the bubble motion needs to be clarified
(especially at the end of the collapse, where the bubble motion
is significantly compressible).

In this paper, we have modified Eq'n. (\ref{eq2}) considering the
effects of compressibility at the bubble interface for both the
liquid and the gas. The modified equation has new terms resulted
from the effects of two coefficients of viscosity of the liquid
and the gas. This work shows that the liquid compressibility
effects in the bubble dynamics originate from two sources. First,
they originate from the liquid motion around the bubble, which
has already been presented in the generalized Eq'n. (\ref{eq1}).
Second, they are initiated through the liquid compressibility
consideration at the bubble interface as presented in this work.
We have performed numerical analysis to investigate the influence
of the modification of Eq'n. (\ref{eq2}).

\section{Compressible Bubble Boundary Equation}

To derive the compressible bubble boundary equation, we assume
that, the motions of the bubble interface and the surrounding
liquid are always spherically symmetric. The continuity equation
and the radial component of the stress tensor can be written as:
\begin{equation}
\label{eq3} \frac{1}{\rho }\left[ {\frac{\partial \rho }{\partial
t}+u\frac{\partial \rho }{\partial r}} \right] = - \frac{\partial
u}{\partial r}-\frac{2u}{r}=-\Delta,
\end{equation}
\begin{equation}
\label{eq4} T _{rr}=-p+\lambda \Delta+2\mu\left(\frac{\partial u
}{\partial r}\right).
\end{equation}
\noindent where, $\rho$, $u$, $p$, and $\Delta $ are density,
velocity, pressure, and divergence of the velocity, respectively.
Also, $\lambda$ is the second coefficient of viscosity. Inserting
$\partial u/\partial r$ from Eq'n. (\ref{eq3}), into Eq'n.
(\ref{eq4}) yields:
\begin{eqnarray}
\label{eq5}T_{rr} &=& -p+(\lambda+2\mu)\left(\frac{\partial u
}{\partial r}+\frac{2u}{r}\right)-4\frac{\mu u}{r}\nonumber \\
&=& -p+(\lambda+2\mu)\triangle-4\frac{\mu u}{r}.
\end{eqnarray}
\noindent The velocity divergence, $\triangle$, can be written as:
\begin{equation}
\label{eq6}
\triangle=-\frac{1}{\rho}\frac{d\rho}{dt}=-\frac{1}{\rho
c^{2}}\frac{dp}{dt},
\end{equation}
\noindent where, the sound speed, $c$, is defined as
$c^{2}=dp/d\rho$. The boundary continuity requirement at the
bubble interface is:
\begin{eqnarray}
\label{eq7} T_{rr}(liquid)\mid_{R}=
T_{rr}\left(gas\right)\mid_{R}+2\frac{\sigma}{R}.
\end{eqnarray}
\noindent Applying Eq'n. (\ref{eq5}) for the gas and the liquid
parts of Eq'n. (\ref{eq7}) leads to:
\begin{eqnarray}
\label{eq7.5} P_{l}+4\frac{\mu\dot{R}}{R} &-&
\left(\lambda+2\mu\right)\triangle_{l}= P_{g}+
4\frac{\mu_{g}\dot{R}}{R}
 \nonumber \\ &-&
\left(\lambda_{g}+2 \mu_{g}\right)\triangle_{g}-2\frac{\sigma}{R},
\end{eqnarray}
\noindent where, $\mu_g$ and $\lambda_g$ are the first and the
second coefficients of viscosity of the gas at the bubble
interface, respectively. Also, $\Delta_{l}$ and $\Delta_{g}$ are
the divergence of velocity of the liquid and the gas,
respectively. Substituting the divergence of velocity for the
liquid and the gas from Eq'n (\ref{eq6}) into Eq'n. (\ref{eq7.5})
yields:
\begin{eqnarray}
\label{eq8} P_{l}+4\frac{\mu\dot{R}}{R} &+&
\left(\frac{\lambda+2\mu}{\rho C^{2}}\frac{dP_{l}}{dt}\right)=
P_{g}+ 4\frac{\mu_{g}\dot{R}}{R}
 \nonumber \\ &+&
\left(\frac{\lambda_{g}+2 \mu_{g}}{\rho_{g}}\frac{d
\rho_{g}}{dt}\right)-2\frac{\sigma}{R},
\end{eqnarray}
\noindent where $\rho_g$ is the gas density at the bubble
interface. Equation (\ref{eq8}) represents the bubble boundary
equation containing all effects of the compressibility and the
viscosity of both the liquid and the gas. Comparison of Eq'ns.
(\ref{eq2}) and (\ref{eq8}) indicates the existence of three new
terms in Eq'n. (\ref{eq8}) due to the liquid and the gas
compressibility and viscosity effects. Here, we concentrate on
the effects of the new term arising from the liquid
compressibility. Therefore, the gas viscosity effects are
neglected due to their smallness relative to the liquid
viscosity as in previous works
\cite{Rayleigh:1917,Noltingk:1951,Herring:1941,Trilling:1952,
Gilmore:1952,Keller:1956,Flynn:1975,Lastman:1979,fstedt:1993,
Nigmatulin:2000,Hickling:1963,Fujikawa:1980,Yasui:1997,Prosperetti:1987,Keller:1980}.
Under this circumstance, Eq'n. (\ref{eq8}) becomes:
\begin{equation}
\label{eq9} P_{l}+\left(\frac{\lambda+2\mu}{\rho
C^{2}}\frac{dP_{l}}{dt}\right)=P_{g}-4\frac{\mu\dot{R}}{R}-2\frac{\sigma}{R}.
\end{equation}

To generalize the argument, we express Eq'ns. (\ref{eq1}) and
(\ref{eq9}) in dimensionless forms. The dimensionless variables
are:
\begin{eqnarray}
\label{eq10} R^{\ast}=\frac{R}{R_{0}}, &\;&
\dot{R^{\ast}}=\frac{\dot{R}}{C},\; \; \; \; t^{\ast}=\frac{t
C}{R_{0}}, \; \; \; \;P_{l}^{\ast}=\frac{P_{l}}{\rho C^{2}},
 \; \; \; \; \nonumber \\
&& P_{g}^{\ast}=\frac{P_{g}}{\rho C^{2}}, \; \; \; \;
P_a^{\ast}=\frac{P_a}{\rho C^{2}}, \; \; \; \;
\end{eqnarray}
\noindent where, $R_{0}$ is the ambient radius of the bubble.
Substituting the dimensionless variables into Eq'ns. (\ref{eq1})
and (\ref{eq9}) results the dimensionless equations:
\begin{eqnarray}
\label{eq11} \left(1-(\eta+
1)\dot{R}^{\ast}\right)\!\!\!&\!\!\!\!\!\!\!\!&\!\!\!\!\!\!
R^{\ast} \ddot{R^{\ast}}+\frac{3}{2}\left(1-\frac{1}{3}(3\eta
+1)\dot{R}^{\ast}\right)\dot{R}^{\ast 2} \nonumber
\\ &=&\!\left(\!{1+(1-\eta)\dot{R}^{\ast}
}\!\right)\!\!\left(P_{l}^{\ast}-P_{a}^{\ast}-P_{0}^{\ast}\right)
\nonumber
\\&+& R^{\ast}\frac{d}{dt^{\ast}}\!\left(P_{l}^{\ast}-P_a^{\ast}\right),
\end{eqnarray}
\begin{equation}
\label{eq12}
P_{l}^{\ast}=P_{g}^{\ast}-4\frac{\mu^{\ast}\dot{R}^{\ast}}{R^{\ast}}
-2\frac{\sigma^{\ast}}{R^{\ast}}-\left(\lambda^{\ast}+2\mu^{\ast}\right)
\frac{dP_{l}^{\ast}}{dt^{\ast}}.
\end{equation}
\noindent The quantities $\sigma^{\ast}$, $\lambda^{\ast}$, and
$\mu^{\ast}$ are the dimensionless form of surface tension, first, and
second viscosity coefficients of the liquid and are defined as:
$\sigma^{\ast}=\sigma / \rho R_{0} C^{2}$, $\lambda^{\ast}=
\lambda / \rho R_{0} C$, and $\mu^{\ast}=\mu /\rho R_{0} C$,
respectively. These dimensionless quantities are basically the
inverse of Weber Number and Reynolds Number.

It should be mentioned that, although the effects of
compressibility in Eq'n. (\ref{eq11}) are in the first order
approximation, but these effects have been introduced completely
in Eq'n. (\ref{eq12}). To quantify the effects of our
modification, we have performed our calculations within the
framework of Eq'n. (\ref{eq11}), using $\eta=1$. Of course,
similar results are obtained using the other values of $\eta$ in
Eq'n. (\ref{eq11}), because of the similarity of the order of
approximation in all equations corresponding to different values
of $\eta$.

\section{Numerical Analysis}

To quantify the effects of the new term in Eq'n.
(\ref{eq12}), numerical analysis was carried out for the
conditions of Single Bubble Sonoluminescence (SBSL)
\cite{Brenner:2002, Barber:1997}. The driving pressure in its
dimensionless form was
$P_{a}^{\ast}(t)=-P_{a}^{\ast}\sin\left(\omega^{\ast}t^{\ast}\right)$,
where $\omega^{\ast}$ is dimensionless angular frequency and is
defined as $\omega R_{0}/C$. To have a well-posed problem, the
value of the gas pressure at the bubble interface, $P_{g}^{\ast}$,
must be specified. It can be determined, in the most complete
approach, from the numerical solution of the conservation
equations for the bubble interior along with Eq'ns. (\ref{eq11})
and (\ref{eq12}) simultaneously. Also, the bubble content changes
during the bubble evolution because of the chemical reactions at
the end of the collapse. In addition, the mass exchange and the
heat transfer between the bubble and the surrounding liquid affect
the bubble interior.

During the past ten years, several different approaches have been
presented to describe the real state of the gas and its evolution
considering the above-mentioned complexities. These approaches,
which at first assumed inviscid and without consideration of
chemical reactions, heat transfer, and mass exchange
\cite{C.C.WU, Moss:1994}, became gradually more complex by
including dissipating effects of radiative transfer
\cite{Kondict:1995}, heat transfer, and viscous gas dynamics
\cite{Voung:1996, Yuan:1998}. Recent gas dynamic model of Storey
and Szeri \cite{Storey:2000} accounts for all effects of chemical
reactions and water vapor evaporation and condensation.
\begin{figure}[t]
\vskip 2mm
\includegraphics[width=6.0cm,height=3.8cm]{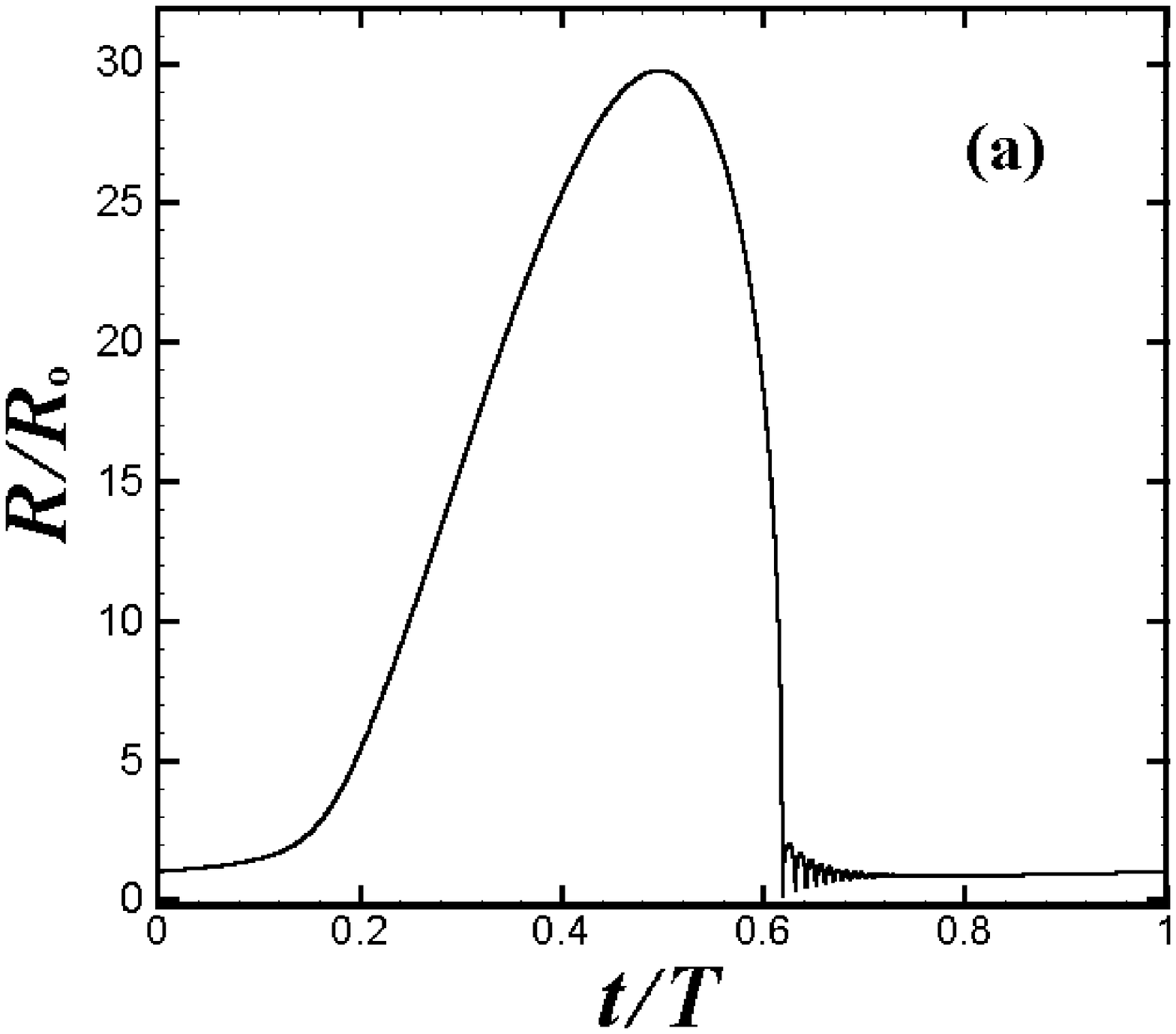}
\vskip 5mm
\includegraphics[width=6.0cm,height=3.8cm]{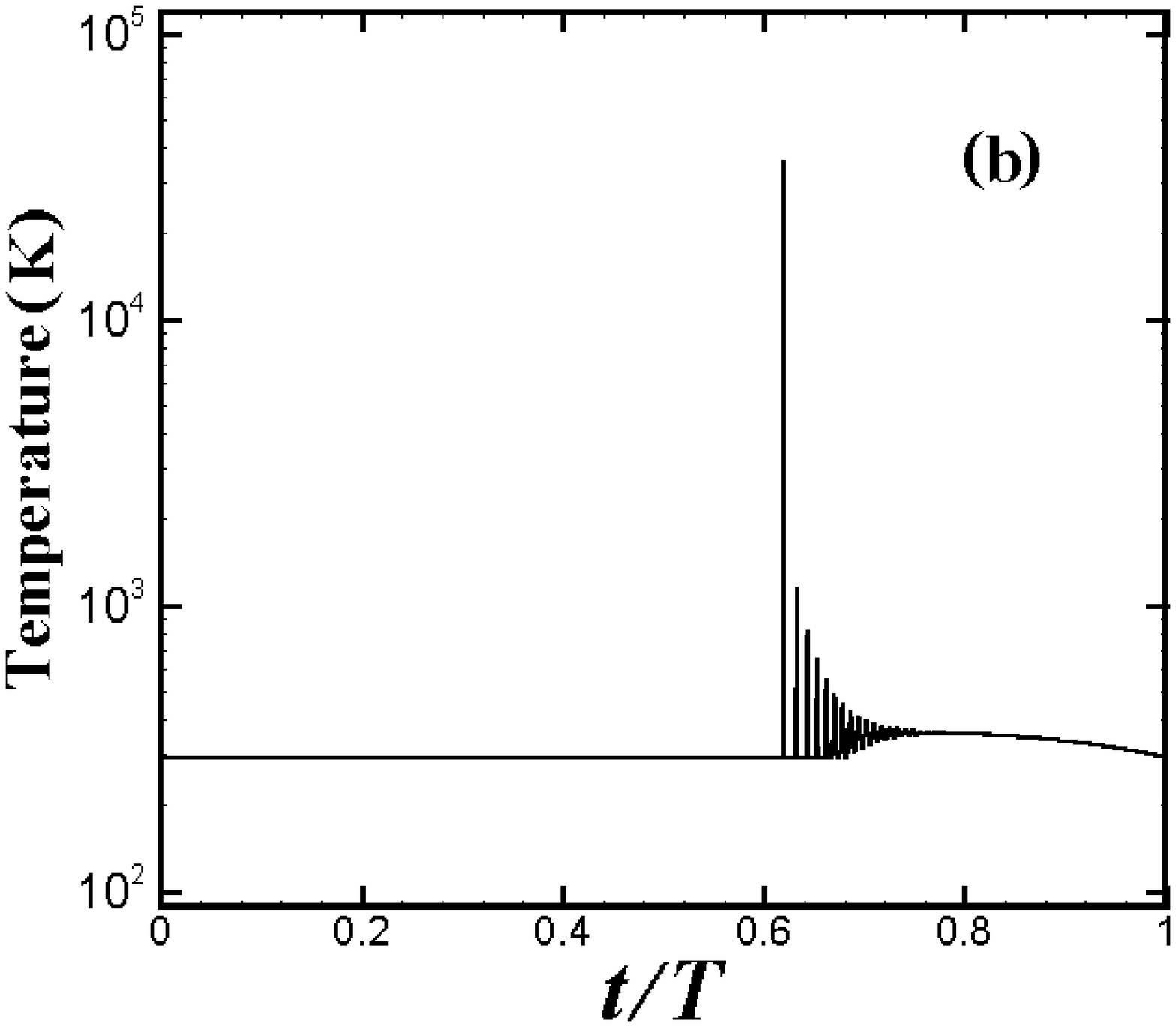}
\caption{Time variations of the bubble characteristics; (a)
radius, (b) gas temperature for the compressible boundary
condition in one period of the applied pressure field.}
\label{fig1:dls}
\end{figure}

During the bubble motion, strong spatial inhomogeneities inside
the bubble are not remarkably revealed unless at the end of the
collapse. Therefore, the uniformity assumption for the bubble
interior seems to be useful and provides many features of the
bubble motion \cite{Barber:1997, Brenner:2002}. By this
assumption, the gas pressure inside the bubble is obtained
assuming polytropic evolution for the gas bubble as:
\begin{equation}
\label{eq29} P_{g}^{\ast}=\left(P_{0}^{\ast}+2\sigma^{\ast}\right)
\left(\frac{1-a^{\ast 3}}{R^{\ast 3}-a^{\ast 3}}\right)^{\gamma},
\end{equation}

\noindent where, Van der Waals equation of state has been used.
Here, $a^{\ast }$ is the dimensionless Van der Waals hard core
radius; $a^{\ast}=a/R_{0}$. The bubble evolution is assumed to be
isothermal ($\gamma=1.0$) for the radii larger than $R_{0}$, when
the bubble moves relatively slow.  While, adiabatic assumption
($\gamma=\Gamma$, where $\Gamma$ is the ratio of the specific
heats of the bubble interior) is applied for the smaller radii,
when the bubble experiences rapid changes \cite{Barber:1997}.

Hilgenfeldt \textit{et al.} \cite{Hilgenfeldt:1999} extended this
model considering $\gamma$ to be a function of the Peclet number
of the bubble. Of course, the Peclet number changes with the
variations of the bubble radius and the bubble velocity. They
showed that, many features of SBSL can be explained by their
model. However, in practice there is very small difference (less
than 5\%) between the Hilgenfeldt \textit{et al.}'s model and the
isothermal-adiabatic model \cite{Putterman:2001}. Although, these
two models can not illustrate production of discontinuities and
shock waves at the end of the collapse, but they are really
useful for our purpose to investigate the importance of the new
term during the collapse. It should be mentioned that,
these models have been used in several recent works in
sonoluminescence studies \cite{ISO-ADI-model}.

We have used isothermal-adiabatic model in this paper. Also, we
have argued about the extension of the results for more
complicated models. Under these circumstances, the time variations
of the bubble properties under the framework of Eq'n (\ref{eq11})
(for $\eta=1$) have been numerically calculated for two cases:
(a) compressible boundary condition (Eq'n. \ref{eq12}) and (b)
incompressible boundary condition (Eq'n. \ref{eq2}). We used
Runge-Kutta method for our numerical analysis. The constants and
the parameters used in the calculations were set for an air bubble
in water at room temperature, $T_{0}=293.0~K$, and atmosphere
pressure, $P_{0}=1.0 ~atm$, \cite{CRC:1991}; $\rho=998.0
~kg/m^{3}$, $C=1483.0 ~m/s$, $\mu=1.01\times10^{-3} ~kg/ms$,
$a^{\ast}=1.0/8.745$, $\sigma=0.0728 ~kgs^{-2}$, and
$\Gamma=1.4$. The second coefficient of viscosity of water was
set to be $\lambda=2.23~\mu$. It was derived from the value of
the bulk viscosity of water at room temperature that,
$\mu_{b}=\lambda+(2\mu/3)=2.9 \mu$ \cite{Pierce:1991}. Also, the
angular frequency of the deriving pressure was
$\omega=2\pi\times26.5~kHz$.
\begin{figure}[t]
\vskip 2mm
\includegraphics[width=6.0cm,height=3.8cm]{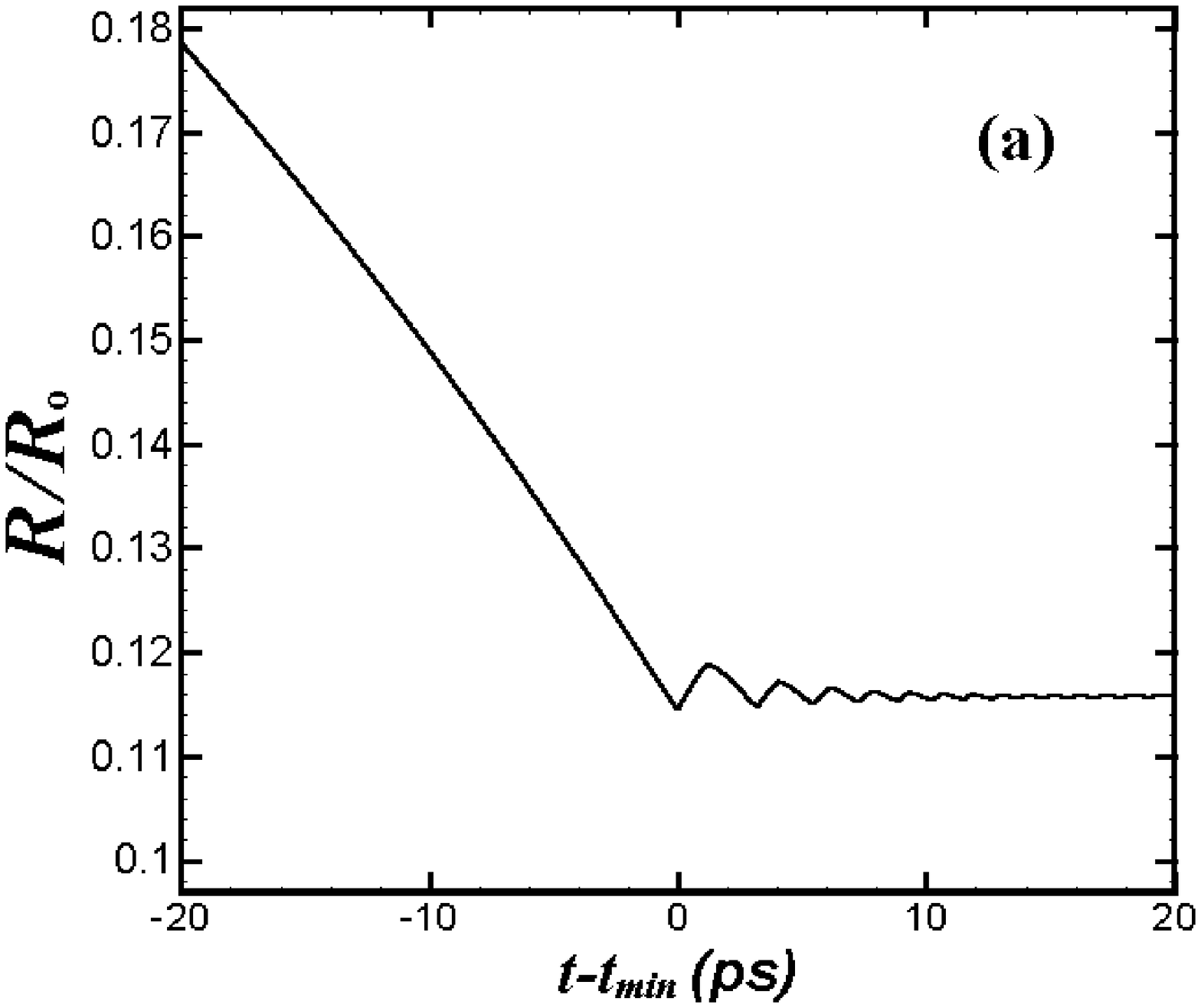}
\vskip 5mm
\includegraphics[width=6.0cm,height=3.8cm]{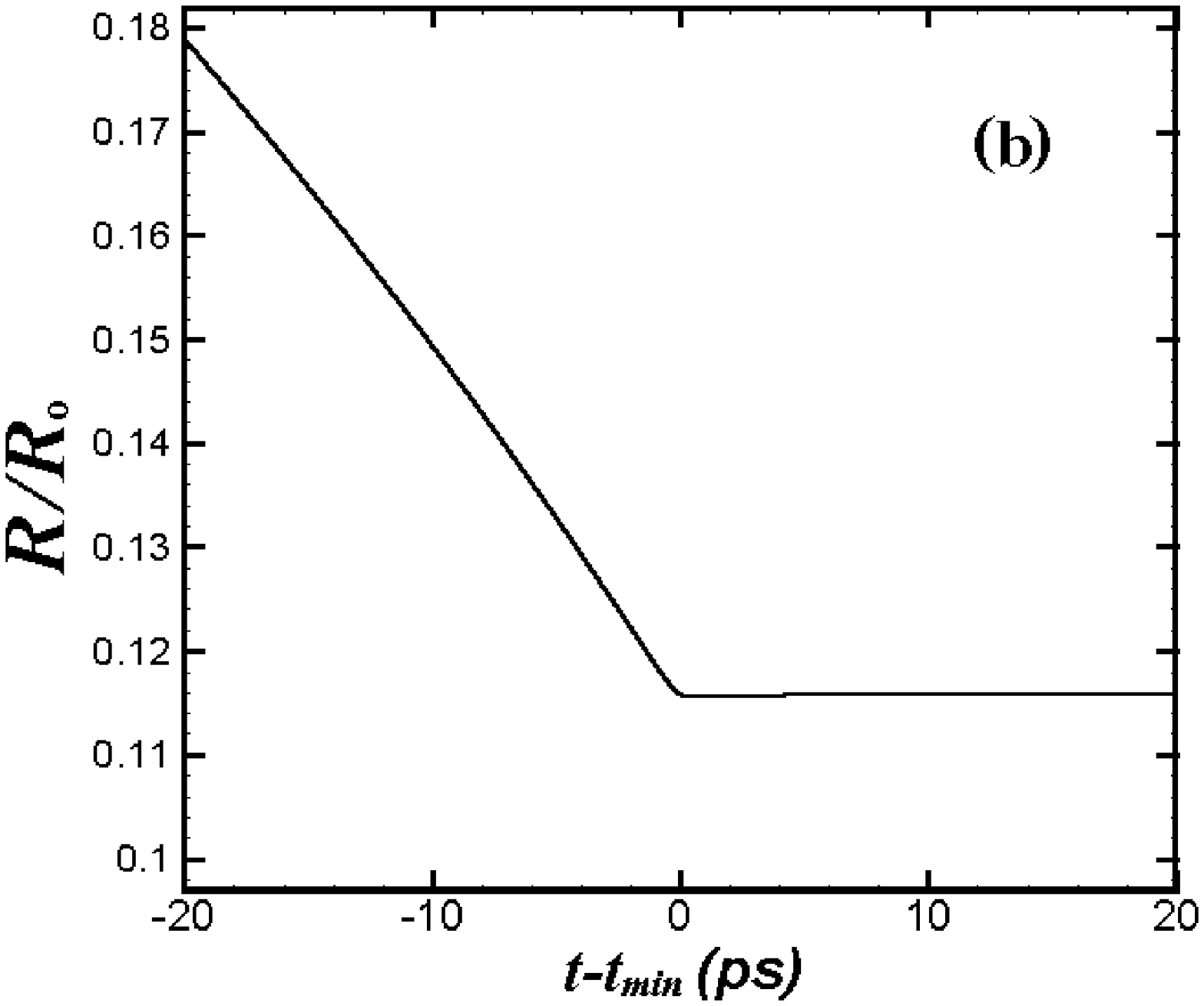}
\caption{Time variations of the bubble radius according to (a)
compressible and (b) incompressible cases. The time interval of
this figures is \textit{40 ps} around the minimum radius
time $(t_{min})$.} \label{fig2:dls}
\end{figure}
\begin{figure}[t]
\vskip 2mm
\includegraphics[width=6.0cm,height=3.8cm]{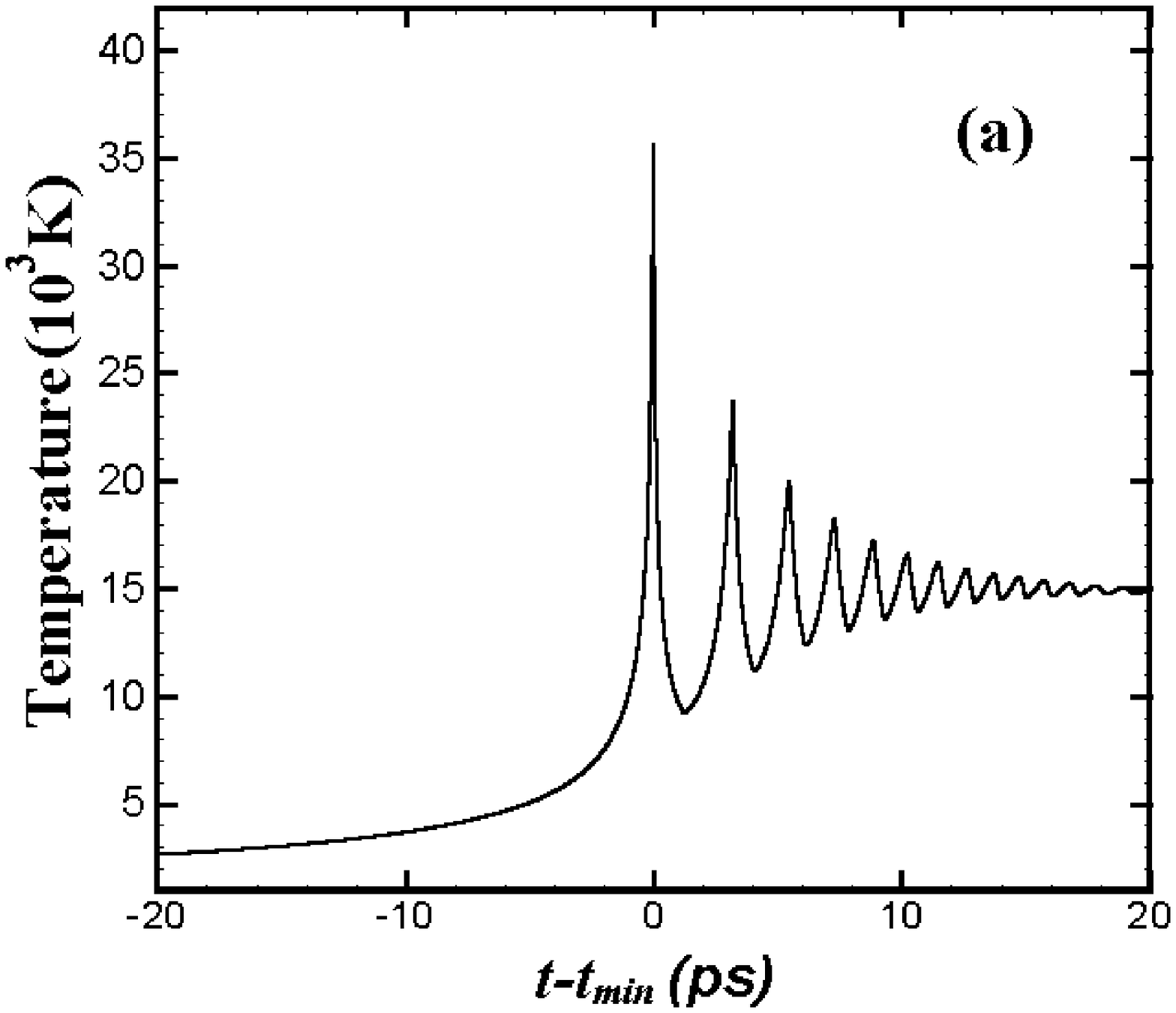}
\vskip 5mm
\includegraphics[width=6.0cm,height=3.8cm]{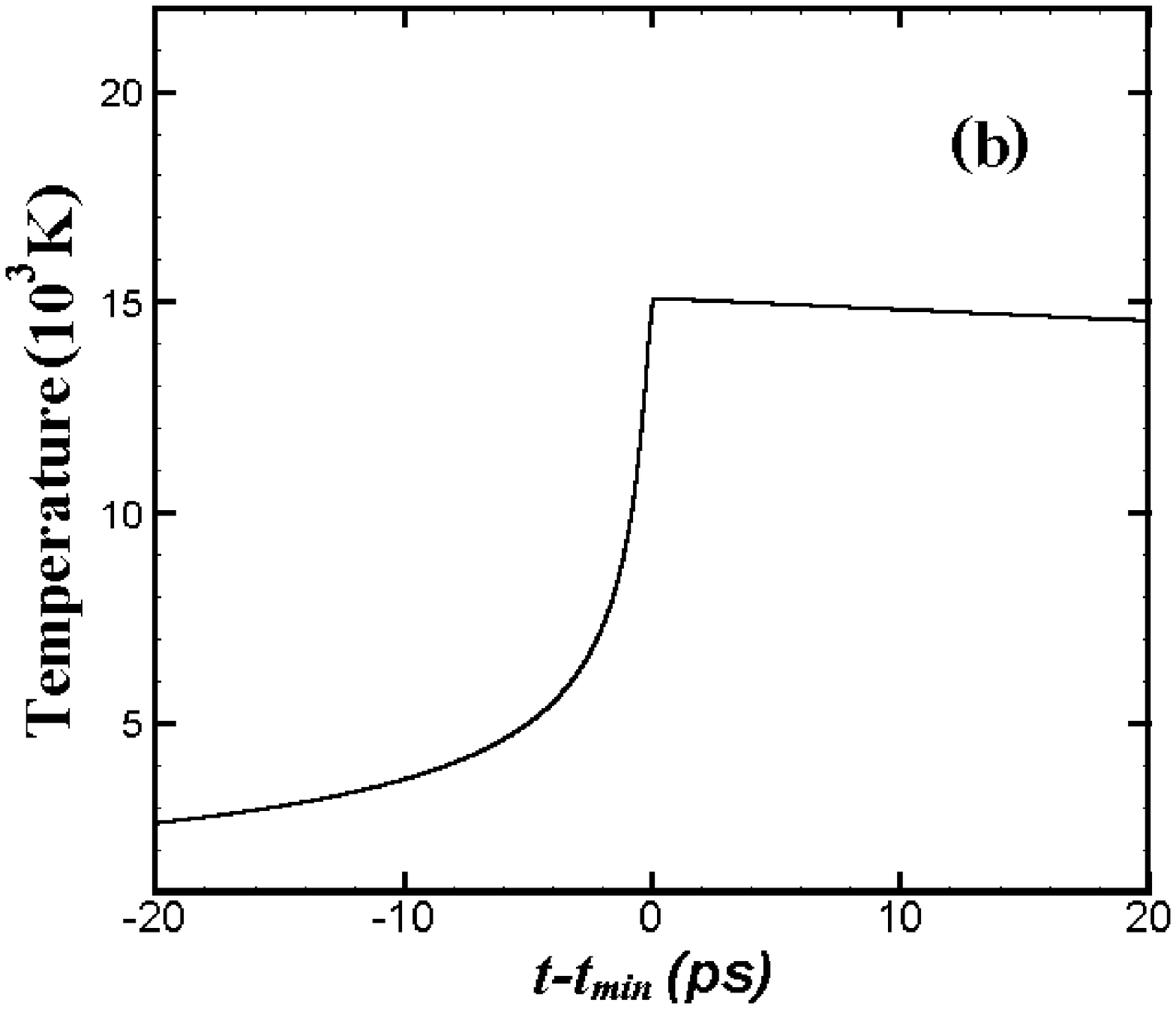}
\caption{Time variations of the gas temperature when the bubble
reaches its minimum radius according to (a) compressible and (b)
incompressible cases.} \label{fig3:dls}
\end{figure}

\section{Results}
The results of our calculations have been illustrated in Figs.
(1-5) for $P_{a}=1.6 ~atm$ and $R_{0}=2.0 ~\mu m$. Similar values
for these parameters were reported in the results of recent experimental works of
Ketterling and Apfel \cite{Ketterling:2000}, Simon \textit{et
al.} \cite{Simon:2001}, and Vazquez \textit{et al.}
\cite{Vazquez:2002}. Figure (1) shows the variations of the
bubble radius and the gas temperature in one period of the
applied pressure field using the compressible boundary condition.
The bubble motion has incompressible characteristics during a
period, except for an infinitesimal time interval at the end of
the collapse. Basically, the new term in Eq'n.
(\ref{eq12}) originates from the compressibility of the liquid
motion. Therefore, it is expected that, its effects not to be
revealed, until the bubble motion becomes significantly
compressible. The results of our calculations clearly confirm this
point. Except for the end of the collapse, the differences between
the bubble characteristics resulting from the compressible and
incompressible boundary conditions are less than $1.0\%$. This
result thoroughly justifies the elimination of the new viscous
term in Eq'n. (\ref{eq12}) for all times, but not for the end of
the collapse. Indeed, in Fig. (1), the results for the bubble
properties of the two boundary conditions completely coincide.
However, the maximum bubble temperature in Fig. 1(b) has a
considerable increase relative to that of the incompressible
case. This discrepancy instigated us for further concentration of
the bubble properties around the minimum bubble radius.
\begin{figure}[t]
\vskip 2mm
\includegraphics[width=6.0cm,height=3.8cm]{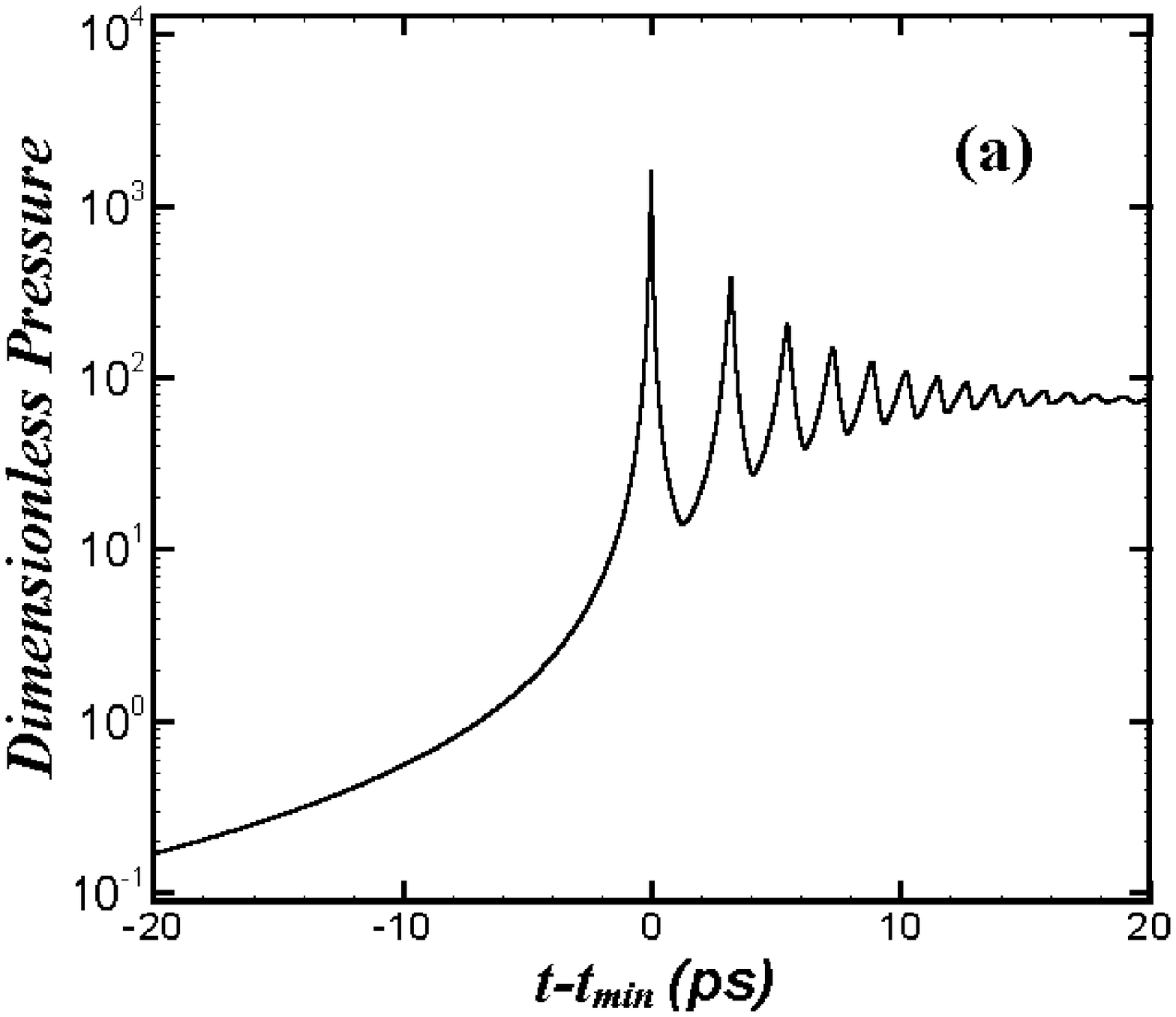}
\\\vskip 5mm
\includegraphics[width=6.0cm,height=3.8cm]{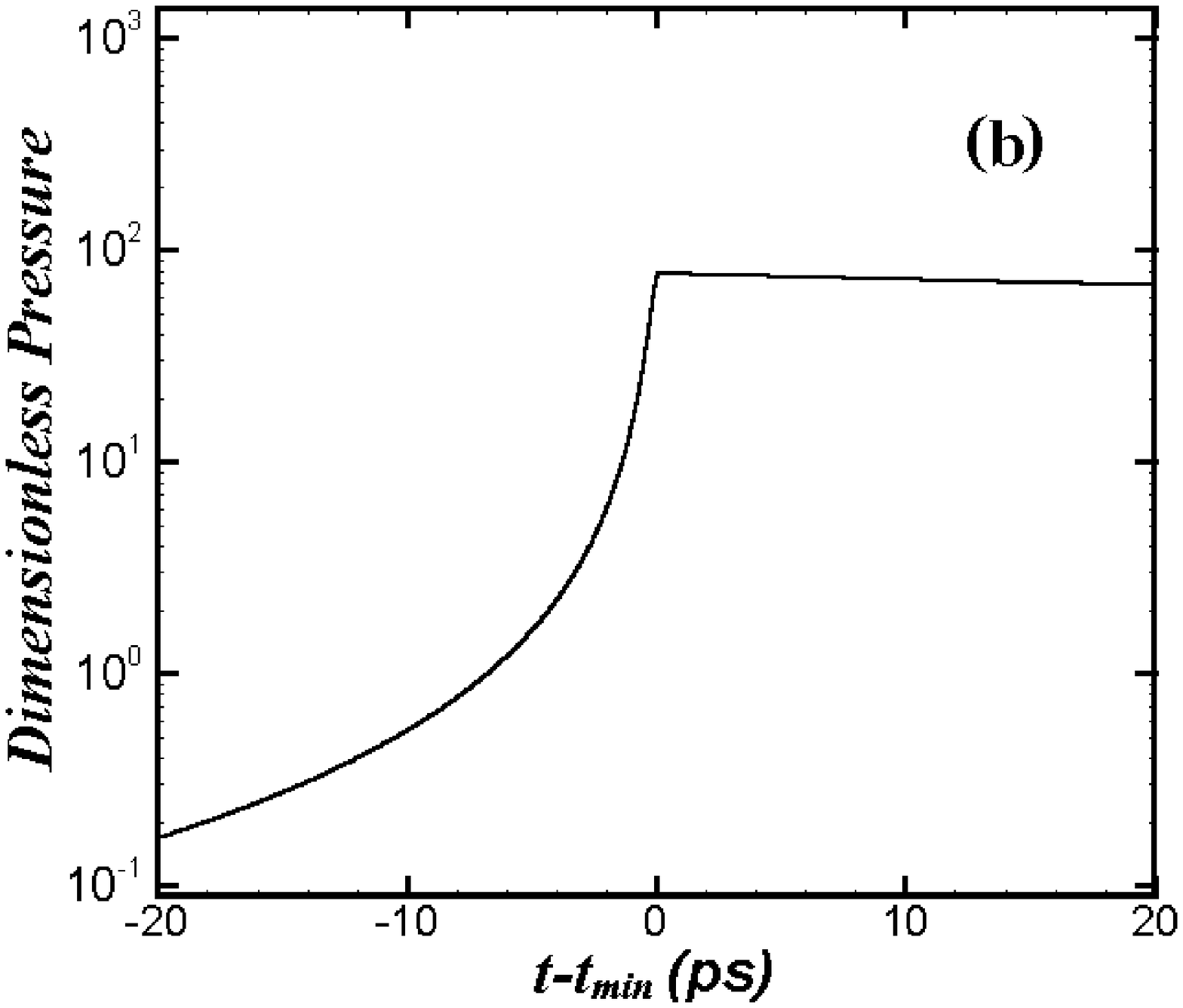}
\caption{Time variations of the dimensionless gas pressure ($P_g /
\rho C^{2}$) around the time of minimum radius according to (a)
compressible and (b) incompressible boundary conditions.}
\label{fig4:dls}
\end{figure}

In Figs. (2-4), we have presented the evolution of the bubble
characteristics using compressible boundary condition (a), and
incompressible boundary condition (b), around the minimum radius
time. The time interval of these figures is \textit{40 ps}.
Since, the bubble experiences its maximum compression at the end
of the collapse, the effects of the new term in Eq'n.
(\ref{eq12}) are more significant during this time interval.
\begin{figure}[t]
\vskip 2mm
\includegraphics[width=6.0cm,height=3.8cm]{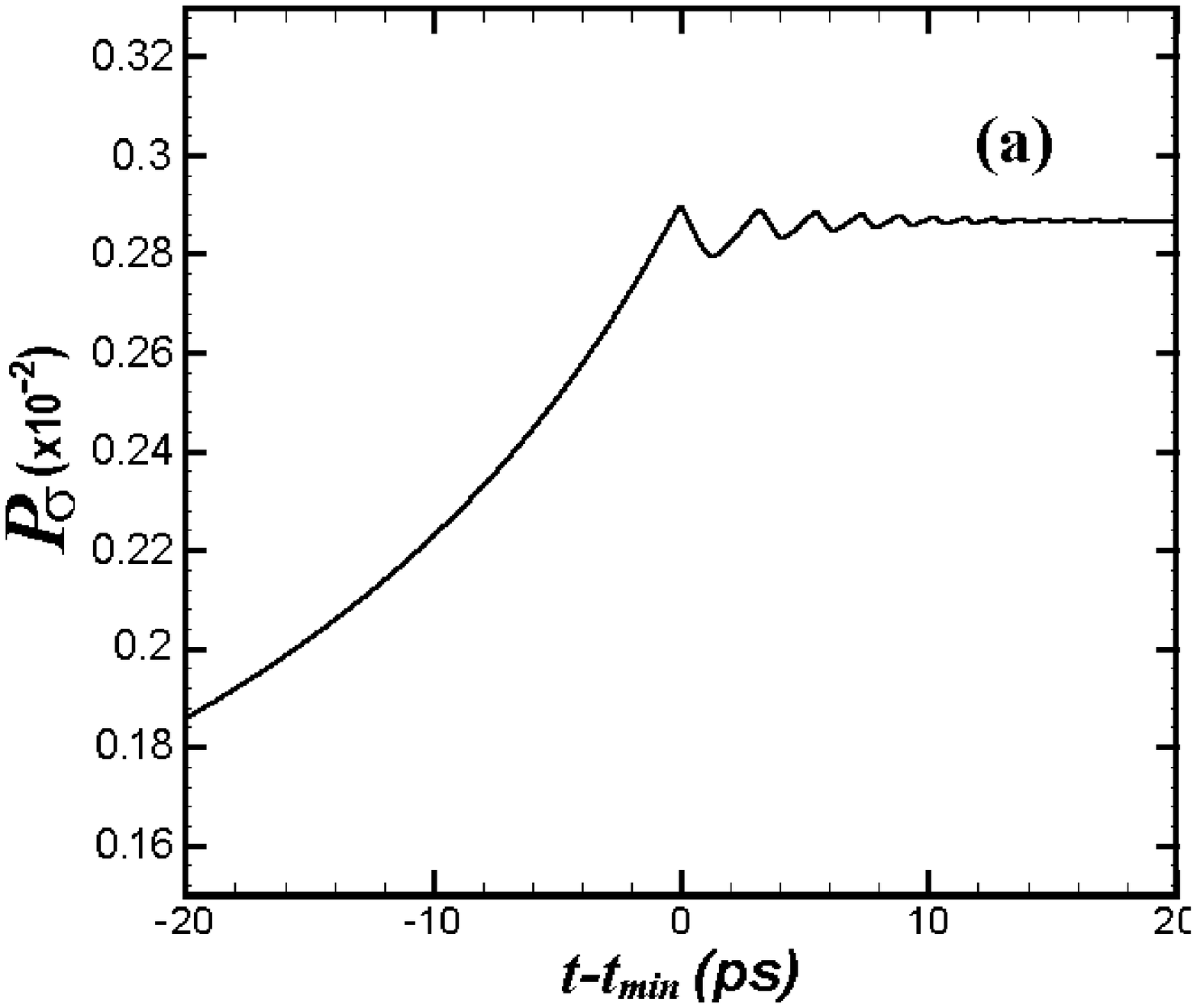}
\vskip 5mm
\includegraphics[width=6.0cm,height=3.8cm]{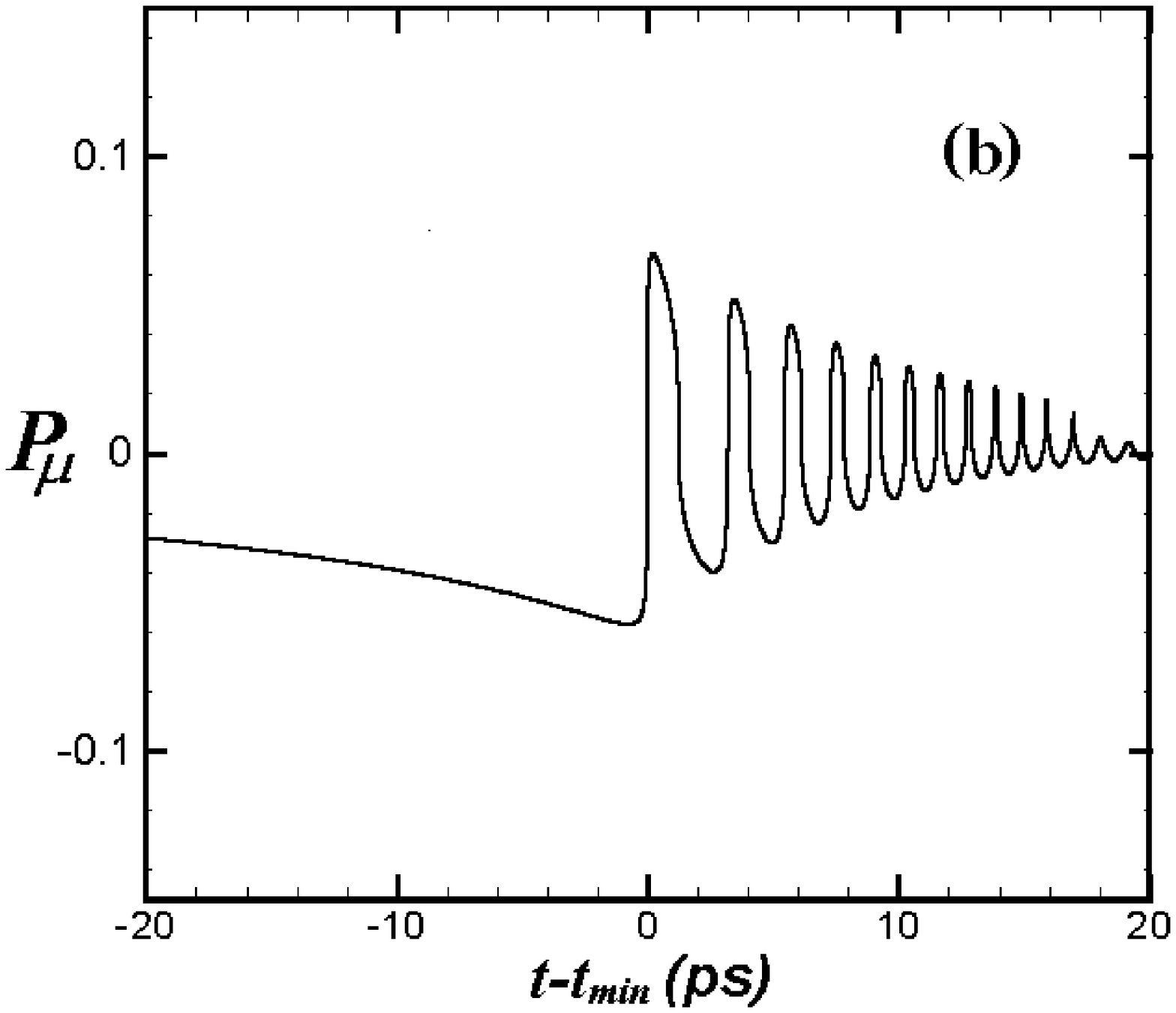}
\vskip 5mm
\includegraphics[width=6.0cm,height=3.8cm]{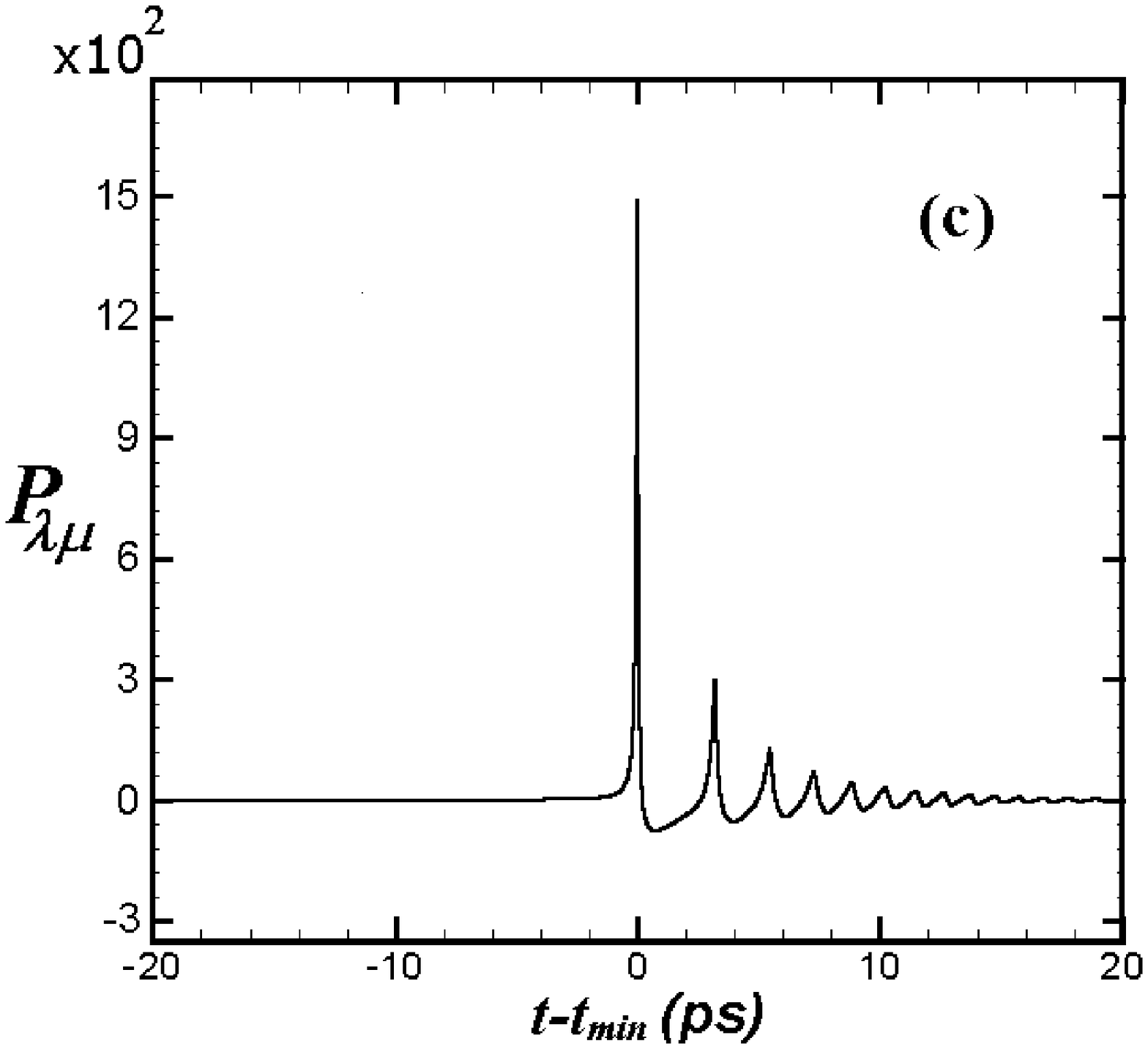}
\caption{Time variations of three dimensionless pressure terms in
equation (\ref{eq12}) namely; (a) surface tension term:
$P_{\sigma}=2 \sigma^{\ast} /R^{\ast}$, (b) damping viscous term:
$P_{\mu}=4\mu^{\ast}\dot{R}^{\ast}/R^{\ast}$, and (c) new viscous
term: $P_{\lambda
\mu}=(\lambda^{\ast}+2\mu^{\ast})(dP_{l}^{\ast}/dt^{\ast})$, when
the bubble reaches its minimum radius.} \label{fig5:dls}
\end{figure}

Figure (2) shows the bubble radius evolution for the two boundary
conditions. This figure illustrates a considerable difference
between the compressible and incompressible cases. After the bubble reaches its
minimum radius, a number of small bouncing oscillations appear in
the graph of the compressible case, which do not occur in that of
the incompressible case. The period of these oscillations is
nearly \textit{3 ps}. The details of our calculations show that,
the times of minimum radius for the two cases are the same. While,
the minimum radius for the new Eq'n. (\ref{eq12}) is 1\% less
than that of the old Eq'n. (\ref{eq2}).

Figure (3) represents the time variations of the gas temperature
near the minimum radius time. It evidently represents that,
introducing the new viscous term in Eq'n. (\ref{eq12}) strongly
affects the gas temperature evolution at the end of the collapse.
After the minimum radius time, the incompressible case illustrates
smooth behavior for the bubble temperature. While, remarkable
sharp peaks appear in the temperature evolution using the
compressible boundary condition. Also, as the bubble goes away
from the minimum radius, the peaks become weaker. Note that, the
value of the maximum temperature increases considerably (nearly
2.4 times) for the new boundary equation.

Comparison of the gas pressure evolution for the two cases is
shown in Fig. (4). Similar behaviors as in Fig. (3) are observed in
this figure for the difference between the new and the old
boundary equations. Note that, the gas pressure is much more
sensitive to the presence of the new viscous term compared to the
gas temperature. In fact, the presence of the new term increases
the maximum pressure more than one order of magnitude.
\begin{figure}[t]
\vskip 2mm
\includegraphics[width=6.0cm,height=3.8cm]{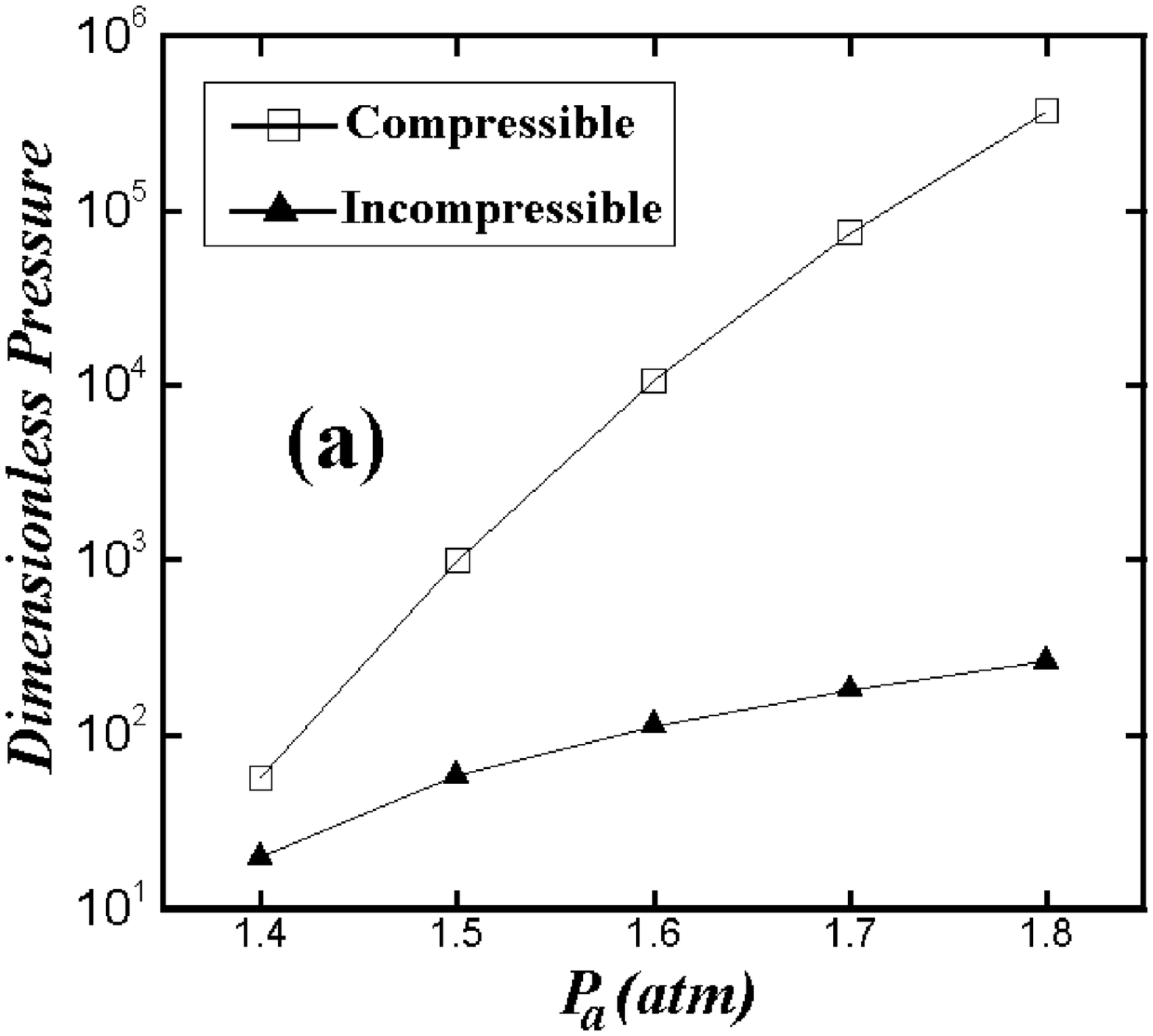}
\vskip 5mm
\includegraphics[width=6.0cm,height=3.8cm]{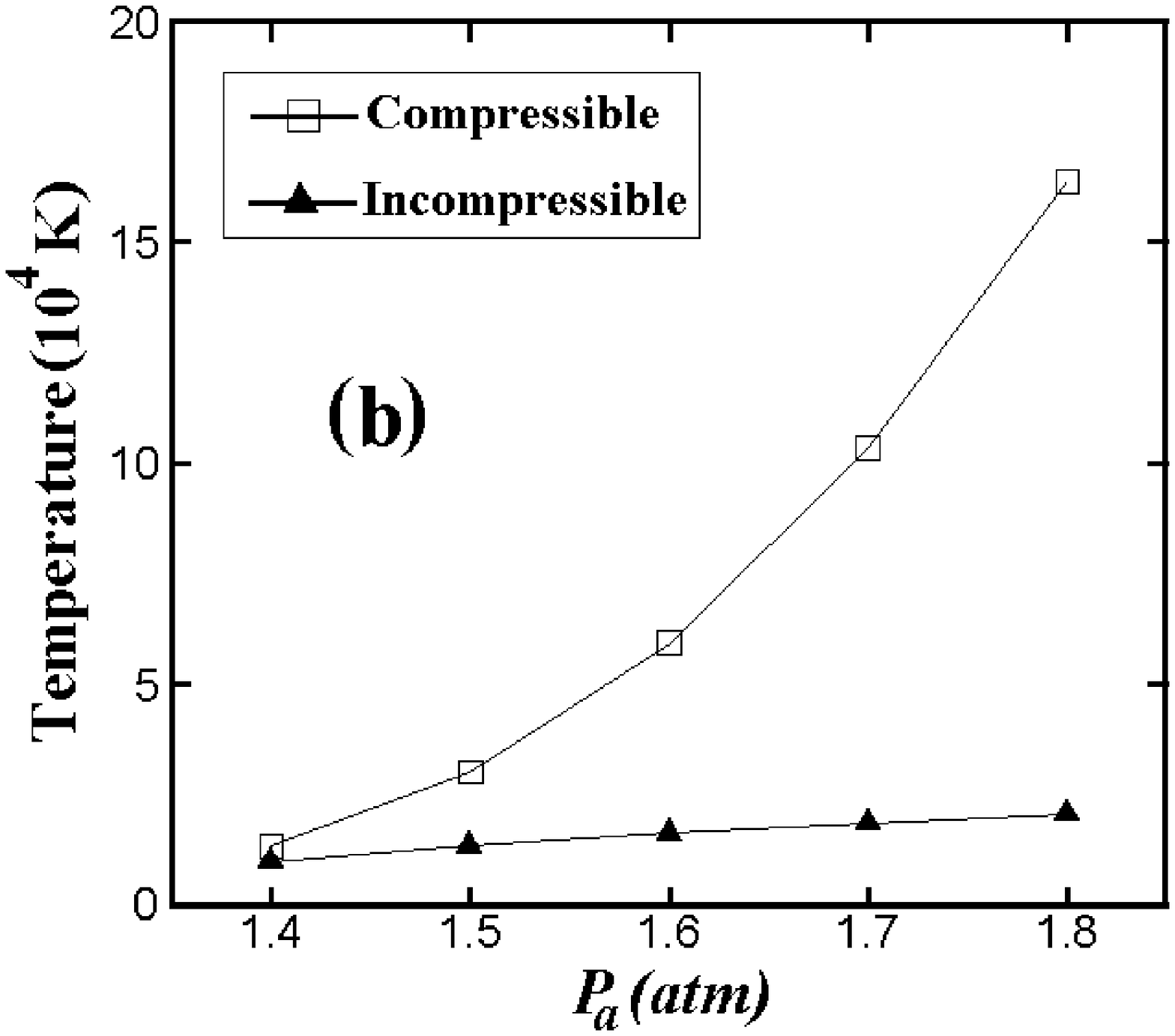}
\caption{(a) The maximum dimensionless gas pressure versus
different amplitudes of the pressure field according to the
compressible (squares) and incompressible (filled triangles)
cases. (b) The maximum gas temperature corresponding to the same
data points of the case (a). The calculated points are for
constant ambient radius $R_{0}=1.5 ~\mu m $. Other constants and
parameters were set as Fig.1} \label{fig6:dls}
\end{figure}

In Fig. (5), the time variations of the three different pressure
terms in Eq'n. (\ref{eq12}) have been illustrated near the minimum
radius. These terms are due to the effects of surface tension and
viscosity, i.e. $P_{\sigma}=2 \sigma^{\ast} /R^{\ast}$,
$P_{\mu}=4\mu^{\ast}\dot{R}^{\ast}/R^{\ast}$, and $P_{\lambda
\mu}=(\lambda^{\ast}+2\mu^{\ast})(dP_{l}^{\ast}/dt^{\ast})$. This
figure clearly shows that, at the end of collapse, the collective
effects of the viscous terms are by far greater than the surface
tension term. Moreover, the new viscous term, $P_{\lambda \mu}$,
is the dominant term in this time interval. Note that, the order
of the maximum values of the three pressure terms are completely
different. $P_{\lambda \mu}$ is the highest (up to $10^{3}$) and
$P_{\sigma}$ is the lowest (less than $10^{-2}$). These results
emphasize that, the elimination of $P_{\lambda \mu}$ is not
reasonable, when the bubble evolves near the minimum radius.

Figures (6) and (7) illustrate the dependence of the peak
pressure and the peak temperature of the bubble to the driving
pressure and to the ambient radius, for the two boundary
equations. Figure (6) is for the case in which, the ambient radius
is constant ($R_{0}=1.5 ~\mu m$), while the amplitude of the
driving pressure is increased from $P_{a}=1.4 ~atm$ to $1.8
~atm$. On the other hand, Fig. (7) shows the results for the
constant amplitude of the driving pressure, ($P_{a}=1.7 ~atm$),
with varying $R_{0}$ from $1.0 ~\mu m$ to $5.0 ~\mu m$. Different
values of $R_{0}$ can be experimentally adapted for a specific
value of $P_{a}$, adjusting the concentration of the dissolved
gas in the liquid \cite{Barber:1997}.
\begin{figure}[t]
\vskip 2mm
\includegraphics[width=6.0cm,height=3.8cm]{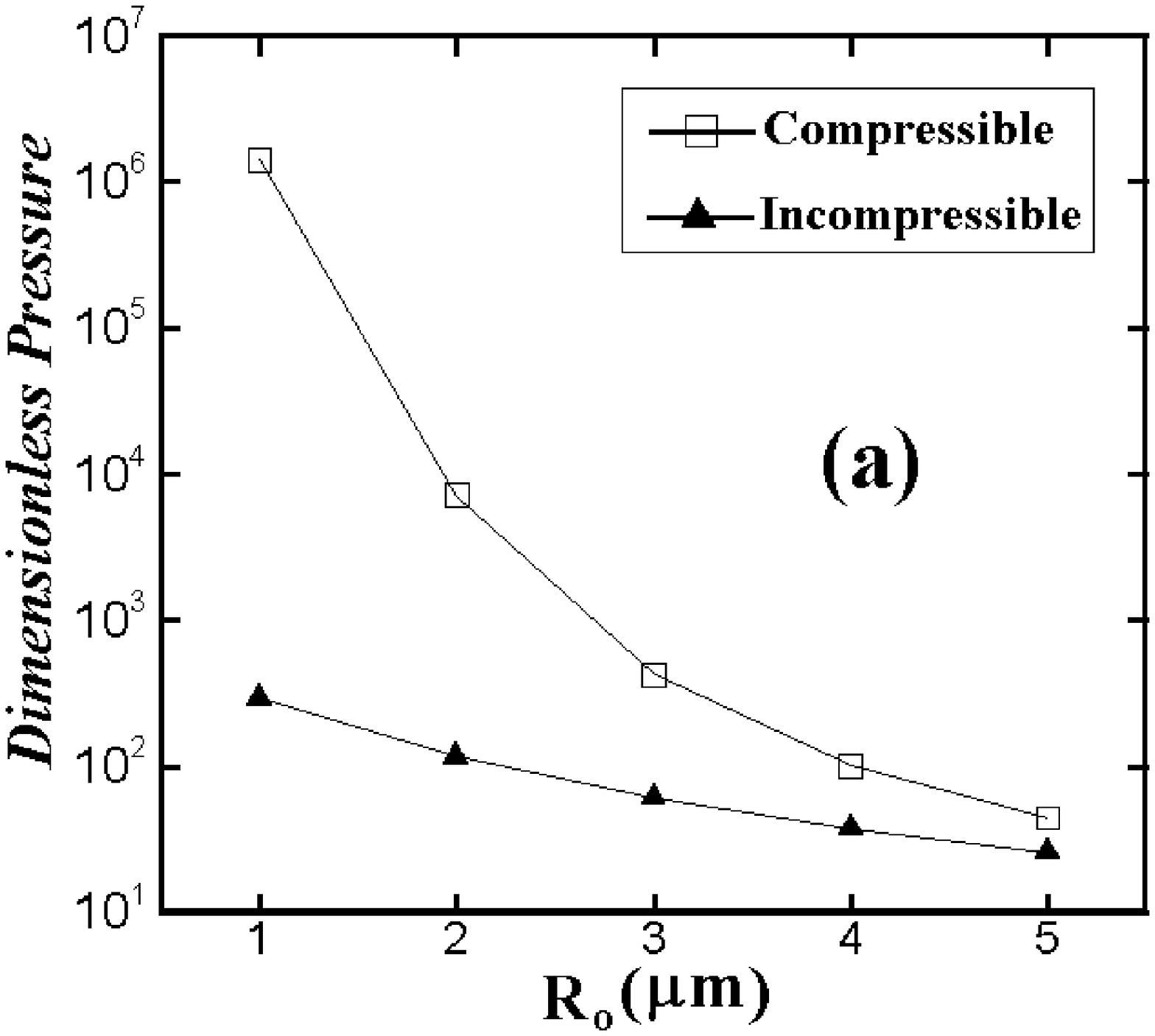}
\vskip 5mm
\includegraphics[width=6.0cm,height=3.8cm]{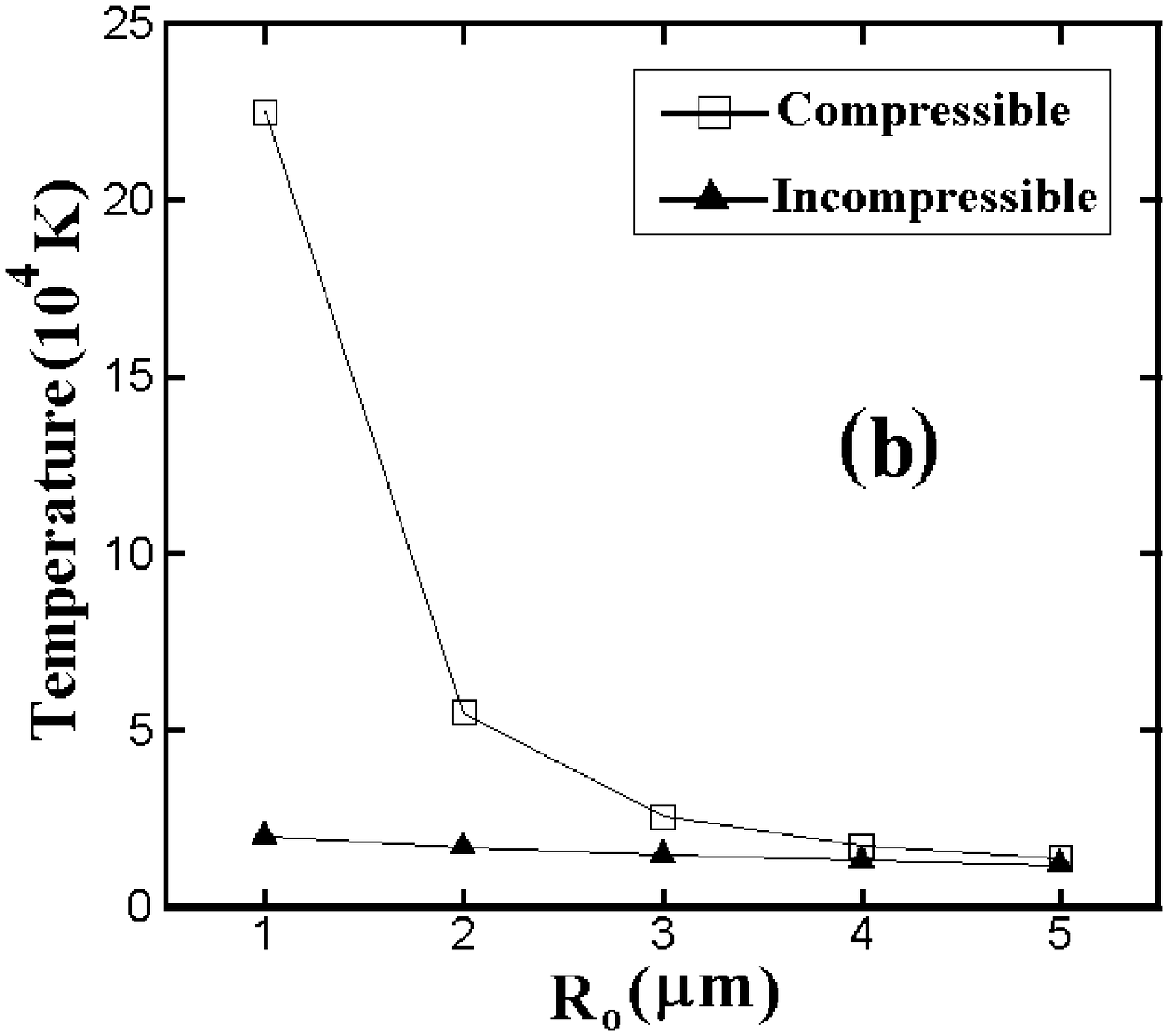}
\caption{(a) The maximum dimensionless gas pressure versus
different bubble ambient radii according to the compressible
(squares) and incompressible (filled triangles) cases. (b) The
maximum gas temperature corresponding to the same data points of
the case (a). The calculated points are for constant driving
pressure amplitude $P_{a}=1.7 ~atm $. Other constants and
parameters were set as Fig.1} \label{fig7:dls}
\end{figure}

The dependence of the maximum pressure and the maximum temperature
of the bubble to $P_{a}$ is represented in Fig. (6). There are
significant differences between these bubble properties resulted
from the new boundary condition with those of the old one. The
differences are more considerable for the higher driving
pressures. Moreover, the differences are more remarkable for the
maximum pressure than the maximum temperature. Note that, for
$P_{a}=1.8 ~atm$, the maximum temperature using the new equations
is about 8 times greater than that of the old ones. While, the
increase of the maximum pressure in this case is about 1400 times.
These results indicate that, the compressibility effects should be
much higher than what is considered in the incompressible case.

The effects of the variation of $R_{0}$ on the mentioned bubble
properties are illustrated in Fig. (7). Similar considerable
differences between the new and the old cases as in Fig. (6) are
also present in this figure. The differences are more significant
for the smaller ambient radii.

Time grid resolution study of the problem shows that, the values
of the bubble properties at the end of collapse are sensitive to
the step size. But, this sensitivity disappears as the resolution
is reduced sufficiently. Figure (8), presents the step size
dependence of the value of maximum gas pressure for the
compressible case, where $P_{a}=1.8 ~atm $ and $R_{0}=1.5 ~\mu
m$. It shows a diminishing step size dependence bellow
$10^{-17}s$.

It should be mentioned that, although the results of the last
figures were acquired by assumption that $\Gamma=1.4$ (the bubble
content was assumed to be a diatomic gas), however the appearance
of the new bubble behavior is independent of the selected value of
$\Gamma$ (and gas content). Figure (9) shows the evolution of the
bubble temperature for two different states; $\Gamma=1.67$
(monoatomic gas) and $\Gamma=1.33$ (polyatomic gas). The
comparison of Figs. (3) and (9) shows that, the new bubble
behavior will be established even when the gas content changes.
Of course, as Figs. (3) and (9) indicates, the configuration and
the number of the new peaks depend on the value of $\Gamma$.
The number of peaks is more for the smaller $\Gamma$. While,
the values of the peak temperatures are greater for higher
$\Gamma$, due to the decrease of the number of the degrees of
freedom.
\begin{figure}[t]
\vskip 2mm
\includegraphics[width=6.0cm,height=3.8cm]{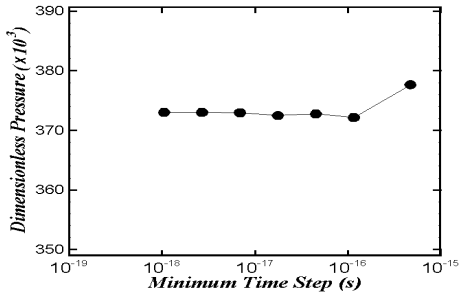}
\caption{Time grid resolution dependence of the maximum gas
pressure for the compressible case, where $P_{a}=1.8 ~atm$ and
$R_{0}=1.5 ~\mu m$.} \label{fig8:dls}
\end{figure}

\begin{figure}[t]
\vskip 2mm
\includegraphics[width=6.0cm,height=3.8cm]{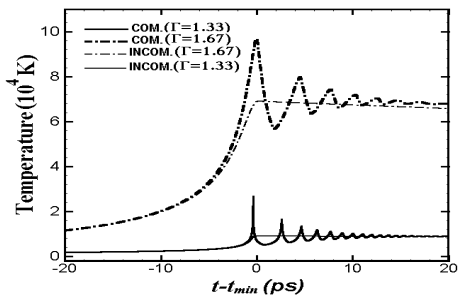}
\caption{Dependence of the gas temperature evolution near the
minimum radius to the value of $\Gamma$, according to
compressible (thick) and incompressible (thin) cases. The dashed
and solid curves are related to $\Gamma=1.33$ and $\Gamma=1.67$,
respectively. Other constants and
parameters are the same as Fig. 1. Note that $t_{min}$ for
$\Gamma=1.33$ is nearly half picosecond less than that of
$\Gamma=1.67$.} \label{fig9:dls}
\end{figure}

\section{Discussion}
Although, the isothermal-adiabatic model used in this study does
not account for the effects of gas dynamics, chemical reactions
and water vapor exchange, but the consideration of these effects
can not cover the importance of the new term at the end of the
collapse. In this section, the influences of these
corrections on the new bubble behavior are argued:

\noindent (i)\textit{ Gas dynamics}: Considerations of the gas
dynamics effects inside the bubble have been presented by several
different approaches \cite{C.C.WU, Moss:1994, Kondict:1995,
Voung:1996, Yuan:1998}. These approaches show that, the overall
bubble temperature and pressure at the collapse in gas dynamics
models are at least no less than those of the isothermal-adiabatic
model. In fact, in the gas dynamics models the bubble experiences
more compression. Therefore, the pressure upon the liquid will be
greater in the gas dynamic models. This means that, the effects of
the new viscous term should even be more remarkable in the gas
dynamics models.

\noindent(ii) \textit{Chemical reactions}: The temperature of the
bubble at the end of the collapse is high enough (higher than
10000 K) to destroy the chemical bonds of N$_{2}$ and O$_{2}$
molecules of the air bubble. The chemical reactions between the
water vapor and the existing oxygen and nitrogen atoms mainly
produce very soluble substances in water (HN, NH3, and HNO3),
which are completely absorbed. Therefore, a sonoluminescence air
bubble mainly contains inert gases. This idea, which was initially
presented by Lohse \textit{et al.} \cite{Lohse:1997} is known as
dissociation hypothesis (DH) and has been confirmed very well by
experimental reports \cite{exp}. The simulation of chemical
reactions at the collapse in a gas dynamics model has been
recently presented by Storey and Szeri \cite{Storey:2000}. Their
work shows that, a considerable decrease appears in the bubble
peak temperature due to the consideration of chemical reactions.
Since most of the reactions at the collapse are endothermic, their
influences can be considered as addition of some extra degrees of
freedom \cite{Brenner:2002}. This means that, the bubble actually
evolves near the minimum radius by an effective exponent,
$\Gamma_{\!e\!f\!f\!}$, which is less than the monoatomic
exponent ($\Gamma=5/3$). However, according to our results, the
appearance of the new behaviors is independent of the values of
$\Gamma$. Therefore, the new bubble behaviors should be
established even when chemical reactions are introduced in the
model.

\noindent(ii) \textit{Water vapor}: Evaporation and condensation
of the water vapor between the bubble and the liquid occur during
the expansion and compression of the bubble. Recent simulations
\cite{Yasui:1997, Storey:2000, Togel:2003} show that, a large
amount of water vapor evaporates into the bubble during the
expansion. Indeed, at maximum radius, about 90\% of the bubble
content is water vapor. During the collapse, the water vapor
molecules condense to the liquid so that, near the minimum radius
only a small fraction of a sonoluminescence bubble is water vapor
and the remaining is mainly inert gases \cite{Yasui:1997,
Storey:2000, Togel:2003}. Presence of these water vapor molecules
inside the bubble decreases the maximum temperature due to the
increase of the bubble's total degrees of freedom. This effect
causes more decrease of the effective exponent
\cite{Brenner:2002}. Therefore, similar to our argument in the last
paragraph, the effects of the water vapor can not also cover the
strong effects of the compressibility consideration of this work.

\section{Conclusions}

The modification of the bubble dynamics equation to account for
the viscosity of a compressible liquid was performed by deriving
a new equation for the bubble boundary. This equation includes a
new term, which has been resulted from simultaneous effects of
viscosity and compressibility of the liquid. The new term is the
prominent term at the end of the collapse, where the bubble is
highly compressed. It exhibits its role by intensifying the
strength of the collapse. The more intense the collapse
is, the more significant the role of the new term is. Moreover,
the new effects can not completely be covered by the
dissipating effects of the water vapor and the chemical reactions.

The results of this work strongly indicate that, the neglect of
the new term at the end of the collapse in the previously derived
equations is not reasonable. Especially, it is more remarkable
for high amplitudes single bubble sonoluminescecne. It is
expected that, these new theoretical results can be confirmed in
the experiments, if resolution of the bubble motion measurement at
the end of an intense collapse is less than 0.1 \textit{ns}. Of
course, a stable high amplitude sonoluminescence bubble can be
produced, if the concentration of the dissolved gas in the liquid
is sufficiently small \cite{Simon:2001}.

\section*{ACKNOWLEDGEMENTS}
This work was supported by Sharif University of Technology and
Bonab Research Center. Partial support of this work by Institute
for Studies in Theoretical Physics and Mathematics is
appreciated. The authors would like to thank Prof. Andrea
Prosperetti for his helpful comments.

\end{document}